\documentclass[aps,preprint,superscriptaddress,showpacs]{revtex4-1}

\usepackage{graphicx}
\usepackage{bm}
\usepackage{amssymb}
\usepackage{color}
\usepackage{tikz}
\usetikzlibrary{plotmarks}
\usepackage{epstopdf}
\begin{document}

\title{Work Fluctuations in a Nonlinear Micromechanical Oscillator Driven Far from Thermal
Equilibrium
}

\author{P. Zhou}
\affiliation{Department of Physics, the Hong Kong University of Science and Technology, Clear Water Bay, Kowloon, Hong Kong, China}

\author{X. Dong}
\affiliation{Department of Physics, the Hong Kong University of Science and Technology, Clear Water Bay, Kowloon, Hong Kong, China}

\author{C. Stambaugh}
\affiliation{Department of Physics, University of Florida, Gainesville, Florida 32611, USA }

\author{H. B. Chan}
\email[]{hochan@ust.hk}
\affiliation{Department of Physics, the Hong Kong University of Science and Technology, Clear Water Bay, Kowloon, Hong Kong, China}

%
\begin{abstract}

We explore fluctuation relations in a periodically driven micromechanical torsional oscillator. In the linear regime where the modulation is weak, we verify that the ratio of the work variance to the mean work is constant, consistent with conventional fluctuation theorems. We then increase the amplitude of the periodic drive so that the response becomes nonlinear and two non-equilibrium oscillation states coexist. Due to interstate transitions, the work variance exhibits a peak at the driving frequency at which the occupation of the two states are equal. Moreover, the work fluctuations depend exponentially on the inverse noise intensity. Our data is consistent with recent theories on systems driven into bistability that predict generic behaviors different from conventional fluctuation theorems.

\end{abstract}

\maketitle

Fluctuation theorems characterize the energy exchange between a system and a reservoir under nonequilibrium conditions \cite{Evans1993,Evans2002}. They are of fundamental interest as fluctuations play an increasing role in small physical and biological systems that are out of thermal equilibrium \cite{Bustamante2005}. In the linear response regime, the probability $P$ of observing the work done $W_{\tau}$ over a time interval $\tau$ by an external force is given by:
\begin{equation}\label{eq:1}
P(W_{\tau})/P(-W_{\tau}) = \exp (W_{\tau} /k_BT)			
\end{equation}
where $k_B$ is the Boltzmann's constant and $T$ is the temperature. In accordance with the second law of thermodynamics, the total entropy tends to increase and the mean value of work done by the external force remains positive. Nonetheless, for individual measurement events there exists a small but finite probability of observing negative work due to fluctuations. 

Experiments on fluctuation theorems have been performed on Brownian particles in optical traps \cite{Carberry2004,Blickle2006,Imparato2007,Ciliberto2013,Gieseler2014}, electrical circuits with an injected current \cite{Garnier2005}, defect centers in diamond \cite{Schuler2005}, quantum dots \cite{Kung2012,Saira2012} and mechanical oscillators \cite{Douarche2006,Ciliberto2013,Bonaldi2009}.  Regardless of the form of the external force and the integration time, the probability distribution of the work done is found to be Gaussian:
\begin{equation}\label{eq:2}
P(W_{\tau}) = (2 \pi \sigma_{\tau}^2)^{-1/2}  \exp [-(W_{\tau}- \langle W_{\tau}  \rangle )^2 /2 \sigma_{\tau}^2] , 		
\end{equation}
which is characterized by the mean work $\langle W_{\tau} \rangle $ and the work variance $\sigma_{\tau}^2=\langle (W_{\tau}-\langle W_{\tau} \rangle )^2 \rangle $. The symmetry function $S(W_{\tau})$ is commonly used to quantify the fluctuation theorem \cite{Douarche2006}:
\begin{equation}\label{eq:3}
S(W_{\tau}) = \ln [P(W_{\tau})/P(-W_{\tau})].
\end{equation}
Since the work distribution is Gaussian, it follows that $S(W_{\tau})  =  \sum{(\tau)} W_{\tau}$, where $\sum(\tau) = 2\langle W_{\tau} \rangle /\sigma_{\tau}^2$. For a linear system modulated by a periodic force, it has been experimentally verified that the ratio of the work variance to the mean work ($\sigma_{\tau}^2/\langle W_{\tau} \rangle$) equals $2 k_B T$ in the steady state, independent of the modulation frequency \cite{Douarche2006}. The above relations are a direct consequence of the linear response theory and the fluctuation dissipation theorem. 

Recently, there has been much interest in fluctuation phenomena in nonlinear systems that are strongly modulated so that multiple nonequilibrium steady states coexist. Examples include Josephson junctions \cite{Siddiqi2004,Lupascu2007}, particles in magneto-optical \cite{Kim2006} and Penning traps \cite{Lapidus1999} and micro- and nano-mechanical oscillators \cite{Aldridge2005,Stambaugh2006a,Chan2007a,Almog2007}. Fluctuations enable the systems to occasionally overcome the activation barrier and switch between the coexisting states \cite{Dykman1979}. In these systems, the relation between work variance and mean work generally cannot be obtained from the linear response theory and the fluctuation dissipation theorem. Theoretical analysis has identified generic features in the work fluctuations of these modulated nonlinear systems that differ from their linear counterparts \cite{Dykman2008}. To our knowledge, fluctuations in nonlinear systems that are periodically driven into multistability have not been experimentally studied in the context of fluctuation theorems.

In this paper, we explore the validity of fluctuation theorems in a periodically driven nonlinear system with two coexisting nonequilibrium steady states. Work fluctuations are measured in a micromechanical torsional oscillator. First, we verify that in the linear regime of small oscillation amplitude, the work fluctuations are consistent with standard fluctuation theorems. Next, we drive the oscillator into bistability. The work distribution remains Gaussian so that $\ln [P(W)/P(-W)] \propto W$. However, the proportionality constant is no longer equal to $1/k_B T$. Near the kinetic phase transition, where the occupation of the two states are comparable, $\sigma^2/\langle W \rangle $ is found to increase by a factor of more than 600 due to fluctuation-induced switching between the two states. Moreover, instead of a linear dependence, $\sigma^2/\langle W \rangle $ varies exponentially with the inverse fluctuation intensity due to the activated nature of switching. These observations are shown to be consistent with predicted generic behaviors for work fluctuations in periodically modulated nonlinear systems \cite{Dykman2008}. 

\begin{figure}
\includegraphics[angle=0]{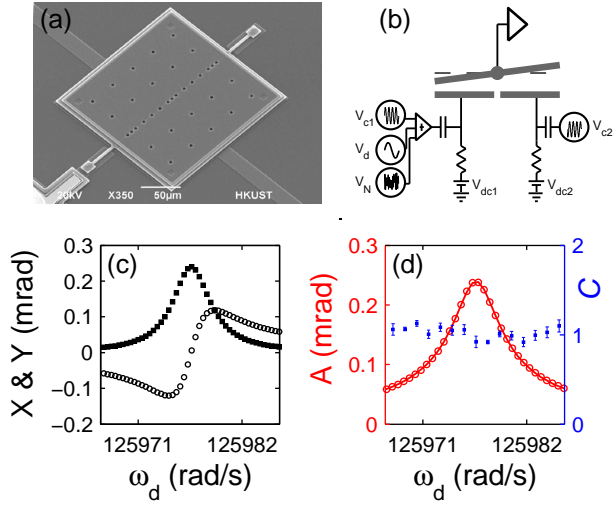} 
\caption{\label{fig:1} (Color online). (a) Scanning electron micrograph of the micromechanical torsional oscillator. (b) Cross-sectional schematic of the device (not to scale) and electrical connections. (c) Oscillation amplitude in phase ($X$, hollow circles) and out of phase ($Y$, solid squares) with the excitation in the linear regime. (d) Hollow circles represent oscillation amplitude $A$ (left axis). Solid squares represent $ C = (\sigma^2/\langle W \rangle)/D$ (right axis).}
\end{figure}

We measure nonequilibrium fluctuations in a periodically driven micromechanical torsional oscillator. The device consists of a 200 $\mu$m by 200 $\mu$m by 3.5 $\mu$m heavily doped polysilicon plate suspended by two torsional springs (Fig.~\ref{fig:1}a) \cite{Stambaugh2006a,Chan2007a}. Two fixed electrodes are located underneath the top plate, enabling us to apply a number of dc, ac and noise voltages to excite and detect rotations of the top plate. Measurement was performed at room temperature and $<  10^{-6}$ Torr. The equation of motion of the plate is given by:
\begin{equation} \label{eq:4}
	 		\ddot{\theta} + 2 \lambda \dot{\theta} + \omega^2_0 \theta + \beta \theta^3 = E/I \cos{(\omega_d t)} + n(t)/I 
\end{equation}
where $\langle  n(t) n(t') \rangle  = 2D \delta(t-t')$. $\theta$ is the angular displacement from the equilibrium position, $\lambda$ ($2.44$ $\mathrm{rad\;s}^{-1}$) is the damping constant, $\omega_0$ ($125972.876$ $\mathrm{rad\;s}^{-1}$) is the resonant frequency, $\beta$ ($-6.0 \times 10^{11} s^{-2})$ is the cubic nonlinear coefficient, $E$ is the amplitude of the periodic driving torque, $I$ ($1.09 \times 10^{-18} $ kg m$^2$) is the moment of inertia, $n(t)$ is the noise torque and $D$ is the noise intensity. For small $\theta$, the nonlinear term $\beta \theta^3$ can be neglected and Eq.~(\ref{eq:4}) reduces to the second order Langevin equation \cite{Landau1976}. In Eq.~(\ref{eq:4}), the periodic torque is produced by a small ac voltage $V_d = V_0 \cos (\omega_d t)$ on top of a much larger dc voltage $V_{dc1}\;(-1.0 V)$ applied to the left electrode, where the driving frequency $\omega_d$ is close to $\omega_0$. The noise torque consists of two components: the torque associated with thermal fluctuations of the top plate and the electrostatic torque due to a noise voltage $V_N(t)$ applied to the left electrode. $V_N(t)$ is Gaussian, with a bandwidth of $1290$ Hz centered at $\omega_0$. Since the bandwidth is much larger than the width of the resonance ($\sim$ 1 Hz), $n(t)$ can be regarded as a white noise torque that increases the effective temperature of the oscillator. 

Motion of the top plate is detected capacitively through induced modulations on a carrier voltage signal. Two ac voltages, $V_{c1}$ and $V_{c2}$, with the same frequency ($f_{c} = 2$ MHz $\gg  \omega_0 / 2 \pi$) and amplitude ($168$ mV) but opposite phase are applied to the two electrodes respectively. Rotation of the top plate leads to changes in the capacitances between the top plate and the two fixed electrodes. As a result, the amplitude of the carrier signal on the top plate is modulated by the plate motion. The voltage on the top plate is measured with a lockin amplifier referenced to $f_{c}$, yielding an output that is proportional to the rotation angle $\theta(t)$. 

In contrast with periodically driven systems in other experiments that studied fluctuation relations \cite{Douarche2006}, our torsional oscillator is strongly underdamped, with a quality factor Q of $\sim$ 30000. The long relaxation time allows us to measure $\theta(t)$ by recording the slowly changing amplitude of the two oscillation quadratures $X(t)$ and $Y(t)$ using a second lockin referenced to $\omega_d$:
\begin{equation}\label{eq:6}
	 	\theta(t) = B [ X(t) \cos (\omega_d t) - Y(t) \sin (\omega_d t )].
\end{equation}
The proportionality constant $B$ is calculated to be $0.02547$ rad V$^{-1}$ based on the device dimensions. With a time constant of $16$ ms ($\ll 1/\lambda$), the measurement uncertainty in $X$ and $Y$ are $\sim 0.094$~$\mu$rad, about 1/3 smaller than the thermal fluctuations of $\sim 0.14$~$\mu$rad with no applied $V_N(t)$. Figure~\ref{fig:1}c shows $X$ and $Y$ measured at a small excitation voltage of $0.0718$ mV as a function of $\omega_d$, in the absence of applied $V_N(t)$. Thermal fluctuations in the motion of the plate are less than $0.1\%$ of the full scale of the plot and the device is well-described by a harmonic oscillator. The hollow circles in Fig.~\ref{fig:1}d represent the amplitude of oscillation $A = \sqrt{(X^2+Y^2)}$.

\begin{figure}
\includegraphics[angle=0]{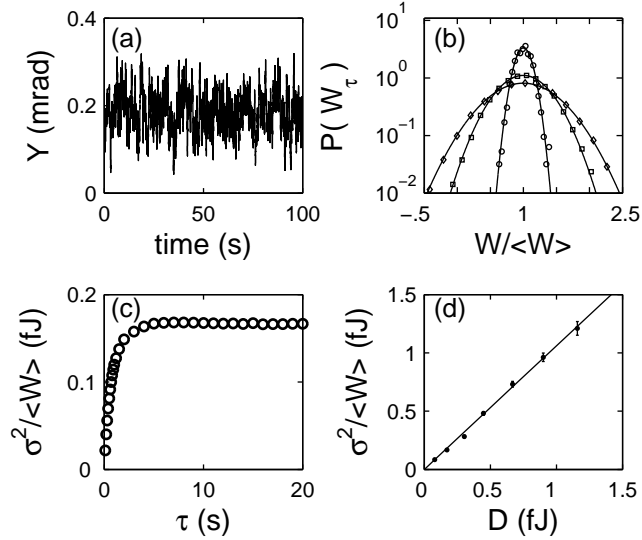} 
\caption{\label{fig:2} (a) $Y(t)$ as a function of time. (b) Probability density function for different $\tau$'s ($\tau = 0.1\,$s ($\diamond$), $1\,$s ($\square$),   $15.0\,$s ($\circ$)) at a fixed $\omega_d$ of 125972.876 $\mathrm{rad\;s}^{-1}$. The lines are Gaussian fits. (c) $\sigma^2/\langle W \rangle $ vs $\tau$ at a fixed $D$ of $0.172$ fJ. (d) $\sigma^2/\langle W \rangle $ vs $D$ at a fixed $\tau$ of 15 s. The line is a linear fit. }
\end{figure}

The work done on the oscillator by the periodic torque $E \cos{(\omega_d t)}$ over time $\tau$ is given by:
\begin{equation}\label{eq:7}
 			W_i(\tau)=\int_{t_i}^{t_i+\tau} dt E \cos(\omega_d t) \dot{\theta}(t)
\end{equation}
where $\tau$ is chosen to be an integer multiple of the period. At the start of observation $t=t_i$, the device has reached the steady state of oscillations. For $\tau\gg1/ \omega_d$, $W_i(\tau)$ can be rewritten in terms of $X(t)$ and $Y(t)$ as:
\begin{equation}\label{eq:8}
W_i(\tau)=\frac{EB}{2}\int_{t_i}^{t_i+\tau} dt [\dot{X}(t)-\omega_d Y(t)]
\end{equation}
where terms that oscillate at frequency $2 \omega_d$ average to zero for slowly varying $X(t)$ and $Y(t)$. In our experiment, the first term in the integrand in Eq.~(\ref{eq:8}) is negligible. Figure~\ref{fig:2}a shows a typical series of $Y$ measured in the presence of noise voltage $V_N$, sampled at a rate of $450$ Hz. 

We verify that the work fluctuations in our oscillator in the linear regime are consistent with the fluctuation theorem. Figure~\ref{fig:2}b shows the probability distribution function $P(W_{\tau})$ for different $\tau$'s. All distributions are well-fitted by Gaussians regardless of $\tau$. In other words, $\ln [P(W_{\tau})/P(-W
_{\tau})] \propto W_{\tau}$, with proportionality constant $2\langle W_{\tau} \rangle / \sigma_{\tau}^2$. For small $\tau$ of $0.1$ s, a number of events with negative work on the oscillator were recorded. As $\tau$ increases, $\langle W \rangle $ is found to increase linearly. In addition, the distribution narrows and negative work occurs too infrequently to be observed. For each $\tau$, the ratio of the work variance to the mean work  $\sigma_{\tau}^2/\langle W_{\tau} \rangle  $ is extracted. As shown in Fig.~\ref{fig:2}c at a fixed $D$ of $0.172$ fJ, $\sigma^2/\langle W \rangle $ increases with $\tau$ up to $\tau \sim 8$ s ($\sim 3/\lambda$) and then approaches a constant value. In the remaining analysis, we choose $\tau = 15$ s to ensure that $\sigma^2/\langle W \rangle $ has approached the saturation value. We also note that for the noise voltages $V_N$ chosen, the noise intensity $D$ is much larger than $k_B T$ ($\sim 4 \times 10^{-21}$ J).

As we described earlier, the conventional fluctuation theorem can be written as $\sigma^2/\langle W \rangle  = CD$.	For a periodically driven linear system, $C$ is equal to one for all driving frequencies. Figure~\ref{fig:2}d plots $\sigma^2/\langle W \rangle $ vs $D$ as the noise $V_N$ injected into the electrostatic torque is increased. $\sigma^2/\langle W \rangle $ is found to be proportional to $D$, with the slope of the linear fit yielding $C$. The measurement of $C$ is repeated for different driving frequencies and plotted as the solid squares in Fig.~\ref{fig:1}d. Our data show that the conventional fluctuation theorem is valid in our device in the linear regime when the oscillation amplitude is small.

  \begin{figure}
\includegraphics[angle=0]{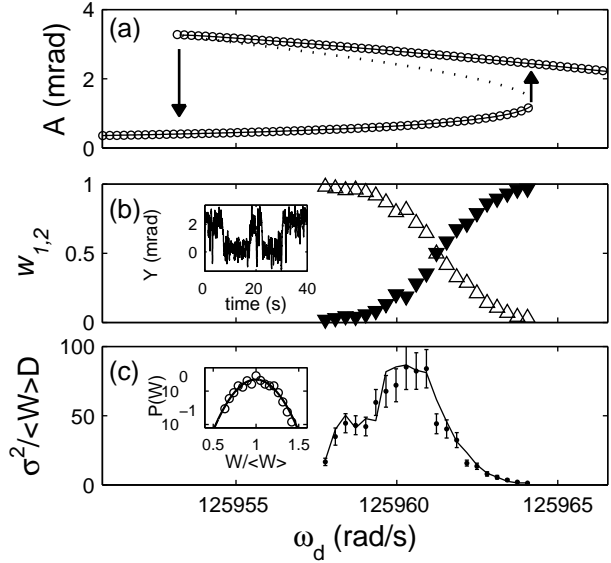} 
\caption{\label{fig:3}(a) Oscillation amplitude versus $\omega_d$. The line is a fit to a damped oscillator with cubic nonlinearity. The arrows indicate jumps in the oscillation amplitude as $\omega_d$ is increased (up arrow) and decreased (down arrow). For frequencies between the two arrows, there are two coexisting stable oscillation states (solid lines) and one unstable state (dotted line). (b) Occupation of the two states vs $\omega_d$. At the kinetic phase transition ($\omega_{kpt} = 125961.232$ $\mathrm{rad\;s}^{-1}$), $w_1$ (inverted solid triangle) and $w_2$ (upright hollow triangle) are almost equal. Inset: $Y(t)$ switches back and forth between two mean values as a function of time. (c) Measured ($\sigma^2/\langle W \rangle)/D$ vs $\omega_d$ (circles) at a fixed $D$ of 9.19 fJ. Inset: $P(W)$ at $\omega_{kpt}$. }
\end{figure}  

Next, we explore work fluctuations in the same oscillator in the nonlinear regime when the cubic term $\beta \theta^3$ in Eq.~(\ref{eq:4}) can no longer be neglected. As the amplitude of the periodic drive is increased, the resonance peak in the frequency response becomes asymmetric and eventually tips over \cite{Landau1976}. Figure~\ref{fig:3}a shows the frequency response of the oscillator, in the absence of the noise voltage $V_N(t)$, for a periodic driving amplitude of $0.95$ mV. For a range of driving frequencies, two non-equilibrium oscillation states with different amplitudes coexist. We denote the oscillation quadratures by $X_j$ and $Y_j$, where $j$ = 1 or 2 refers to the two oscillation states.
When the noise is weak, the system remains in one of the oscillation states $j$, with
$X(t)$ and $Y(t)$ fluctuating randomly about $X_j$ and $Y_j$ respectively. Occasionally a large fluctuation occurs, inducing the oscillator to switch from one oscillation state into the other by overcoming an activation barrier $R$ (inset of Fig.~\ref{fig:3}b) \cite{Dykman1979}. We refer to the former as ``intrastate fluctuations'' and the latter as ``interstate fluctuations''. Since both the amplitude and phase of the two oscillation states differ from each other, the power that they absorb from the periodic driving torque is also significantly different \cite{Dykman2008}. During interstate transitions, the power switches back and forth between the two mean values, producing large fluctuations in work. Similar to systems in thermal equilibrium, $\sigma^2$ is proportional to $\langle W \rangle$ and $P(W)$ remains Gaussian (inset of Fig.~\ref{fig:3}c). However, due to the activated nature of switching, it has been predicted that the dependence of the work fluctuations on the noise intensity and driving frequency differs considerably from the conventional fluctuation theorem \cite{Dykman2008}.

For a bistable oscillator, the rate of noise-activated switching out of state $j$ is given by \cite{Dykman1979}
\begin{equation}\label{eq:9}
\nu_j \propto \exp(-R_j/D)
\end{equation}
where $j =1$ or $2$. It follows that the occupation of the two states can be written as  $w_{1,2}=\nu_{2,1}/(\nu_1+\nu_2)$. $w_{1,2}$ are strongly dependent on the driving frequency. In the bistable region, $R_1$ ($R_2$) monotonically increases (decreases) as the frequency increases. On the low frequency side of the bistable region, $R_2$ exceeds $R_1$ and the oscillator predominantly resides in the low-amplitude state, and vice versa for the high frequency end. There exists a narrow range of intermediate frequencies (around $\omega_{kpt}\sim 125961.232 $~$\mathrm{rad\;s}^{-1}$ in Fig.~\ref{fig:3}b) where $R_1 \sim R_2$ and the occupation of the two states are comparable. In this ``kinetic phase transition'', the system switches back and forth between the two oscillation states, leading to a variety of fluctuation phenomenon \cite{Dykman1990a,Stambaugh2006a,Moon2013}.

Figure~\ref{fig:3}c plots the ratio $\sigma^2/\langle W \rangle D$ as a function of $\omega_d$, at a fixed noise intensity $D$ of $9.19$ fJ. The observation time $\tau$ of $100$ s is chosen to be longer than the mean residence time in the oscillation states. At the kinetic phase transition, the ratio attains a maximum. This strong dependence of $\sigma^2/\langle W \rangle  D$ on frequency is distinct from the frequency independent behavior of the linear regime shown in Fig.~\ref{fig:1}d. The peak value of $\sigma^2/\langle W \rangle  D$ at the kinetic phase transition exceeds the constant $C$ in the linear regime by a factor of about 90.

  \begin{figure}
\includegraphics[angle=0]{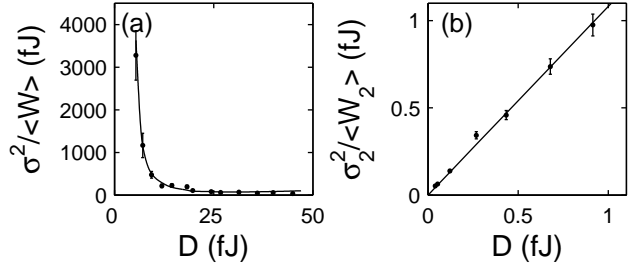} 
\caption{\label{fig:4}Dependence of $\sigma^2/\langle W \rangle $ on noise intensity for (a) interstate fluctuations and (b) intrastate fluctuations. The line in (a) is calculated using Eq.~(\ref{eq:10}). The line in (b) is a linear fit.}
\end{figure}  

Apart from the strong frequency dependence, the work fluctuations in a nonlinear oscillator vary with the noise intensity $D$ in a manner that is different from ordinary equilibrium and modulated linear systems. Figure~\ref{fig:4}a plots $\sigma^2/\langle W \rangle $ vs $D$ at the kinetic phase transition of the nonlinear oscillator. Instead of a linear dependence on $D$ as in a linear system (Fig.~\ref{fig:2}d), $\sigma^2/\langle W \rangle $ decreases as $D$ increases. Such behavior can be understood in terms of fluctuation induced switching between coexisting oscillation states. According to Eq.~(\ref{eq:9}), larger noise leads to an exponential increase in the switching rate. For a fixed observation time $\tau$, frequent interstate transitions enable more effective averaging of the two distinct values of work of the two states, resulting in a reduction in the work variance. 

The work fluctuations near the kinetic phase transition are dominated by interstate transitions. Intrastate fluctuations, although much smaller, can also be measured. By reducing the noise intensity and choosing sufficiently short $\tau$'s, the system has negligible probability of switching out of the state. Figure~\ref{fig:4}b shows the dependence of $\sigma_2^2/\langle W_2 \rangle $ on the noise intensity for intrastate fluctuations in the low amplitude state. Similar to the linear regime in Fig.~\ref{fig:2}d, $\sigma_2^2/\langle W_2 \rangle $ is proportional to $D$, with the proportionality constant measured to be $1.08$. 

We compare our experimental results on work fluctuations with theoretical predictions \cite{Dykman2008}. For $\tau \gg 1/\nu$ in a bistable system, the work variance is given by:
\begin{equation}\label{eq:10}
\sigma^2 = w_1\sigma^2_1 + w_2\sigma^2_2 + \frac{2w_1w_2}{ (\nu_{1}+\nu_{2})\tau}(W_1-W_2)^2.
\end{equation}
 The first two terms represent the intrastate fluctuations weighted by the state occupations. The third term involves interstate fluctuations that dominate the work fluctuations at the kinetic phase transition. In Fig. 3c, the solid line represents the calculated value of the ratio  $\sigma^2/\langle W \rangle D$ using Eq.~(\ref{eq:10}) with the measured values of $w_{1,2}, W_{1,2}, \nu_{1,2}$, $\langle W \rangle $ and $D$. The values of $\sigma_{1,2}$   are linearly extrapolated from measurements at much lower $D$ to avoid interstate transitions. There is good agreement between theory and experiment at all driving frequencies with no fitting parameters. In Fig.~\ref{fig:4}a the solid line is the calculated $\sigma^2/\langle W \rangle $ as a function of $D$ at the kinetic phase transition. As explained earlier, the sharp rise of $\sigma^2/\langle W \rangle $ at decreasing $D$ occurs due to the activated nature of transitions. Using $\nu_1 = \nu_2 \propto \exp(-R/D) $ in Eq.~(\ref{eq:10}) yields the fitted value $R$ of $2.86 \times 10^{-14}$ J. The good agreement between theory and measurement confirms that interstate switching is responsible for the large work fluctuations near the kinetic phase transition.    

Besides the work fluctuations due to interstate transitions at the kinetic phase transition, recent theoretical analysis indicates that intrastate fluctuations in nonlinear modulated systems also leads to deviations of work fluctuations from conventional fluctuation theorems. Specifically, the work variance is expected to diverge near the bifurcation points at the boundary of the bistable frequency range with characteristic scaling exponents \cite{Dykman2008}. Further experiments are warranted to test such predictions and reveal other phenomena of work fluctuations in nonlinear systems where fluctuation-dissipation theorem is not applicable.

We thank M. I. Dykman for useful discussions and K. Ninios for assistance. This work is supported by the Research Grants Council of Hong Kong SAR (Project No. 600312). C.S. was supported by NSF DMR-0645448.



\end{document}